\documentclass[aps,pre,twocolumn,floatfix]{revtex4-1}

\usepackage{amsmath}
\usepackage{amssymb}
\usepackage{graphicx}
\usepackage{sidecap}
\usepackage{wrapfig}
\usepackage{subcaption}
\usepackage{xcolor}
\usepackage{natbib}
\usepackage{dcolumn}
\usepackage{datetime}
\usepackage{fancyhdr}

\newcommand{\Drosophila}{\textit{Drosophila}}
\newcommand{\wherebib}{/home/mholcomb/SHAREDLAB/Common_BIB}

\usepackage{ragged2e}
\usepackage{localdefs}

\DeclareCaptionJustification{justified}{\justifying}
\newlength{\figtotwidth}
\begin{document}


\title{Embryo as an active granular fluid: stress-coordinated cellular
  constriction chains}

\author{Guo-Jie Jason Gao} \affiliation{Department of Mechanical
  Engineering, National Taiwan University}

\author{Michael C. Holcomb} \affiliation{Department of Physics, Texas
  Tech University}

\author{Jeffrey H. Thomas} \affiliation{Department of Cell Biology and
  Biochemistry, Texas Tech University Health Sciences Center}

\author{Jerzy Blawzdziewicz} \affiliation{Department of Mechanical
  Engineering, Texas Tech University}


\begin{abstract}

Mechanical stress plays an intricate role in gene expression in
individual cells and sculpting of developing tissues. However,
systematic methods of studying how mechanical stress and feedback help
to harmonize cellular activities within a tissue have yet to be
developed.  Motivated by our observation of the cellular constriction
chains (CCCs) during the initial phase of ventral furrow formation in
the \textit{Drosophila melanogaster} embryo, we propose an active
granular fluid (AGF) model that provides valuable insights into
cellular coordination in the apical constriction process.  In our
model, cells are treated as circular particles connected by a
predefined force network, and they undergo a random constriction
process in which the particle constriction probability $P$ is a
function of the stress exerted on the particle by its neighbors.  We
find that when $P$ favors tensile stress, constricted particles tend
to form chain-like structures.  In contrast, constricted particles
tend to form compact clusters when $P$ favors compression.  A
remarkable similarity of constricted-particle chains and CCCs observed
\textit{in vivo} provides indirect evidence that tensile-stress
feedback coordinates the apical constriction activity.  Our
particle-based AGF model will be useful in analyzing mechanical
feedback effects in a wide variety of morphogenesis and organogenesis
phenomena.

\end{abstract}

\maketitle

\section{Introduction}
\label{Introduction}

Multicellular organisms utilize mechanical stress fields as a means of
guiding tissue growth, triggering genetic expression and cell
division, and enhancing the robustness of morphogenetic processes
\cite{Mammoto-Ingber:2010,%
  Miller-Davidson:2013,%
  Zhang-Labouesse:2012,%
  Shiu-Weiss-Hoying-Iwamoto-Joung-Quam:2005,%
  Farge:2003,%
  Pouille-Ahmadi-Brunet-Farge:2009,%
  Idema-Liu:2013,%
 Idema-Dubuis-Kang-Manning-Nelson-Lubensky-Liu:2013%
}.
This emerging evidence of the role of mechanical feedback in
orchestrating cellular-level activity has given an impetus to analyze
mechanical processes involved in biological development.

While the existing evidence of the crucial role of mechanical
triggering in sculpting tissues and coordinating cell behavior is
indisputable, quantitative understanding of how cell communication via
long-range mechanical stress fields harmonizes cell activity is
fragmented and incomplete.  We propose that local mechanical
interactions and global stress fields in tissues can be qualitatively
represented and analyzed by modeling the tissue as an active granular
medium.

Tissues are a conglomeration of deformable, discrete objects (cells)
that mechanically interact through direct contact and adhesion, and
they are large enough for thermal motion fluctuations to be
neglected.  Cells, however, are not merely passive, deformable
objects.  They are in fact subject to genetically prescribed active
deformations that can give rise to large-scale cellular flows
resulting in tissue-wide structural changes.  One particularly
striking example of such active cellular flows is the collection of
regional cellular motions (morphogenetic movements) by which an embryo
changes from a single layer of cells around a yolk center into a
triple-layered structure (the process known as gastrulation).

\begin{figure}[b]
  \newcommand{\figwidth}{0.2\figtotwidth}
  \centering 
  \includegraphics{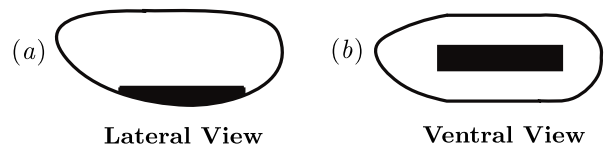}
  \\[15pt] 
  \caption{Schematic of a \Drosophila\ embryo with the mesoderm
    primordium (ventral furrow) region marked by the black stripe.
    \subfig{a} Lateral (side) view; \subfig{b} ventral (bottom) view.
    Cells in the marked region are active and in the unmarked region
    are passive during the initial phase of the apical constriction
    process.}
  \label{schematic}
\end{figure}

Gastrulation occurs in most animals.  The active granular fluid (AGF)
model proposed in this study is motivated by specific features of
gastrulation in the common fruit fly
(\Drosophila\ \textit{melanogaster}).  The first morphogenetic
movement of gastrulation---i.e., ventral furrow formation---is
initiated by the constriction of the outer (apical) faces of the cells
on what will become the underside (ventral side) of the fruit fly (see
the schematic representation in Fig.\ \ref{schematic}).  The
constrictions produce negative spontaneous curvature of the active
region of the cell monolayer, eventually leading to its invagination.

\begin{figure}
  \begin{center}
    \includegraphics{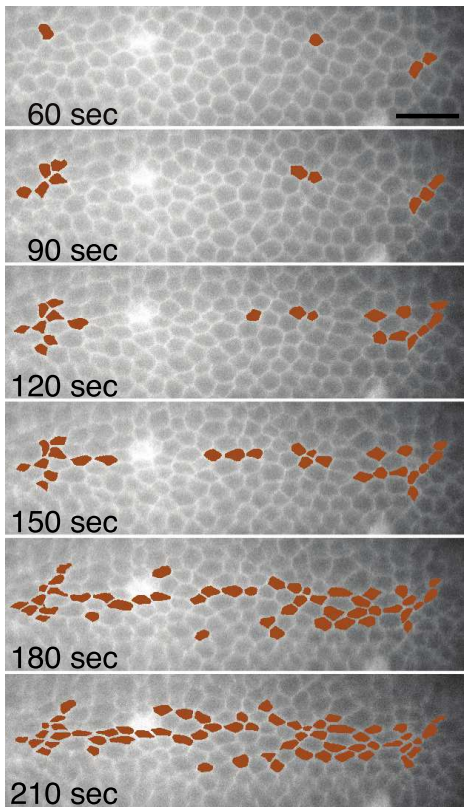}
    \\[-11pt]
    
  \end{center}
  \caption{Time-lapse images of the ventral side of a wild-type
    \Drosophila\ embryo (the region indicated by the black stripe in
    Fig.\ \ref{schematic}) during the slow phase of the
    apical constriction process.  Marked cells (in brown) illustrate
    the propagation of cellular constriction chains (CCCs). Scale
    bar${}=20\mu\textrm{m}$.}
  \label{cellconstchains}
\end{figure}

In the modeling effort presented in this study, we are concerned with
the initial phase of ventral furrow formation, before the actual
tissue invagination occurs.  During this initial constriction
phase, approximately 40\% of the cells gradually constrict in a
seemingly random order.  Furrow formation is subsequently completed
with a rapid, coordinated constriction of the remaining active cells
(fast phase)
\cite{Sweeton-Parks-Costa-Wieschaus:1991}.

While apical constrictions during the initial slower phase of
ventral furrow formation are generally accepted to be an uncorrelated
stochastic process \cite{Sweeton-Parks-Costa-Wieschaus:1991}, we show
here that this phase is not completely random.  A close inspection of
the distribution of constricted cells (see Fig.\ \ref{cellconstchains}
and discussion in Sec.\ \ref{Epithelial Tissue as an Active Granular
  Medium}) reveals the presence of chain-like arrangements.  We call
these arrangements \textit{cellular constriction chains} (CCCs),
because they are remarkably reminiscent of force chains which are
observed in granular media
\cite{Howell-Behringer-Veje:1999,%
Behringer-Howell-Kondic-Tennakoon-Veje:1999}.

We propose that CCCs in the mesoderm primordium of the
\Drosophila\ embryo form as a result of coordination of cell activity
through mechanical stresses.  Namely, constrictions of cells (which
are bonded to the surrounding cells) produce tensile stresses that
propagate along tensile force chains (analogous to compressive force
chains in granular matter).  The presumed coupling of these strongly
correlated stresses to the constriction probability of individual
cells causes formation of chain-like structures of constricted cells.

We note that it was recently reported that a correlation between the
rachetted contractile pulses of constricting cells had been observed
\cite{Xie-Martin:2015}.  However, a robust description of mechanical
interactions and, possibly, coordination between the apically
constricting cells has yet to be formulated.

Our approach draws on ideas developed for granular matter.  Below we
introduce an AGF model to describe collective cell
behavior during the initial phase of apical constrictions in the
ventral furrow region.  We present our proof-of-concept calculations
along with a qualitative comparison with \textit{in vivo} observations.

\section{Epithelial tissue as an active granular fluid}
\label{Epithelial Tissue as an Active Granular Medium}

\subsection{A Brief review of relevant biology}

Gastrulation in \Drosophila\ begins around 3 hours after fertilization and is
completed through multiple morphogenetic movements which are driven by
region-specific cell activities \cite{Leptin:1999}. These regions are
established as a result of a cascading pattern formation caused by
symmetry breaking events which occur during the creation of the egg
(oogenesis)
\cite{StJohnston-NussleinVolhard:1992,%
Riechmann-Ephrussi:2001,%
Huynh-StJohnston:2004,%
Vaneeden-StJohnston:1999}.
The region of cells that undergoes the apical constriction of interest
is known as the mesoderm primordium and is actually internalized by
ventral furrow formation.

The mesoderm primordium is composed of a band of cells on the ventral
side of the embryo which take up approximately 80\% of its length and
20\% of its circumference \cite{Leptin:1999}, as schematically
depicted in Fig.\ \ref{schematic}.  Mesoderm primordial cells are capable of
mechanical activity, due to expression of regulatory genes
\textit{twist} and \textit{snail}
\cite{Leptin-Grunewald:1990,%
Leptin:1991,%
Ip-Maggert-Levine:1994,%
Seher-Narasimha-Vogelsang-Leptin:2007,%
Martin-Kaschube-Wieschaus:2009,%
Pouille-Ahmadi-Brunet-Farge:2009}
established during the preceding phase of embryo patterning.  Cells
outside the mesoderm primordium undergo passive deformations under
applied stresses, but otherwise remain mechanically inactive during
the initial slower stage of ventral furrow formation
\cite{Sweeton-Parks-Costa-Wieschaus:1991}.

\subsection{Force chains vs.\ cellular-constriction chains}

During the slower phase of ventral furrow formation, a growing number
of mesoderm primordium cells undergo apical constriction.  Initially
the constrictions occur at random locations, but as time progresses
the constricted cells tend to form CCCs, correlated chain-like
patterns (see the highlighted cells in time-lapse images in
Fig.\ \ref{cellconstchains}; the experimental details are described in
Appendix \ref{experimental methods}).  Similar CCC-like patterns can
also be discerned in classical images available in the literature (see
Fig.\ 4 of Ref.\ \cite{Sweeton-Parks-Costa-Wieschaus:1991}), but to
our knowledge such structures have not been explicitly reported, and
their significance has not yet been analyzed.

We argue that CCCs occur as a result of coordination of apical
constrictions via mechanical feedback.  As discussed in
Sec.\ \ref{Introduction}, our mechanical feedback conjecture is based
on the close resemblance between CCCs and chains of interparticle forces
that occur in granular media
\cite{Howell-Behringer-Veje:1999,%
Behringer-Howell-Kondic-Tennakoon-Veje:1999}.
In compressed
or strained granular matter, individual force chains consist of a
sequence of pairwise compressive forces between interacting particles
that are jammed together; the  chains act as a path along which the 
stress in the material is propagating. Similar tensile-force chains
occur in systems of bonded particles
\cite{Tordesillas-Tobin-Cil-Alshibli-Behringer:2015}.

Force-chain related structures have  been observed in a variety of
systems including emulsions, foams, and colloidal glasses
\cite{Kondic-Goullet-OHern-Kramar-Mischaikow-Behringer:2012}.
Force chains, resulting from collective interactions between
individual constituent particles, are a prevalent phenomenon in
condensed particular matter and, therefore, we propose that they also
occur in active cell packings constituting a developing tissue.

Epithelial cells in a \Drosophila\ embryo are bonded to their
immediate neighbors through specific protein formations (adherens
junctions).  Thus, both tensile and compressive stresses can be
transmitted in cellular systems.  The distribution of stress through
force chains effectively shields the rest of the material, creating low
stress regions.  Assuming that the constriction
probability for a given cell is affected by the forces exerted on it by
the neighboring cells, such mechanical coupling may result in
a non-random microstructured distribution of the constricted
cells.  In the following sections we examine this possibility using our 
AGF model.

\begin{figure*}
  \centering
  \includegraphics[width=0.93\textwidth]{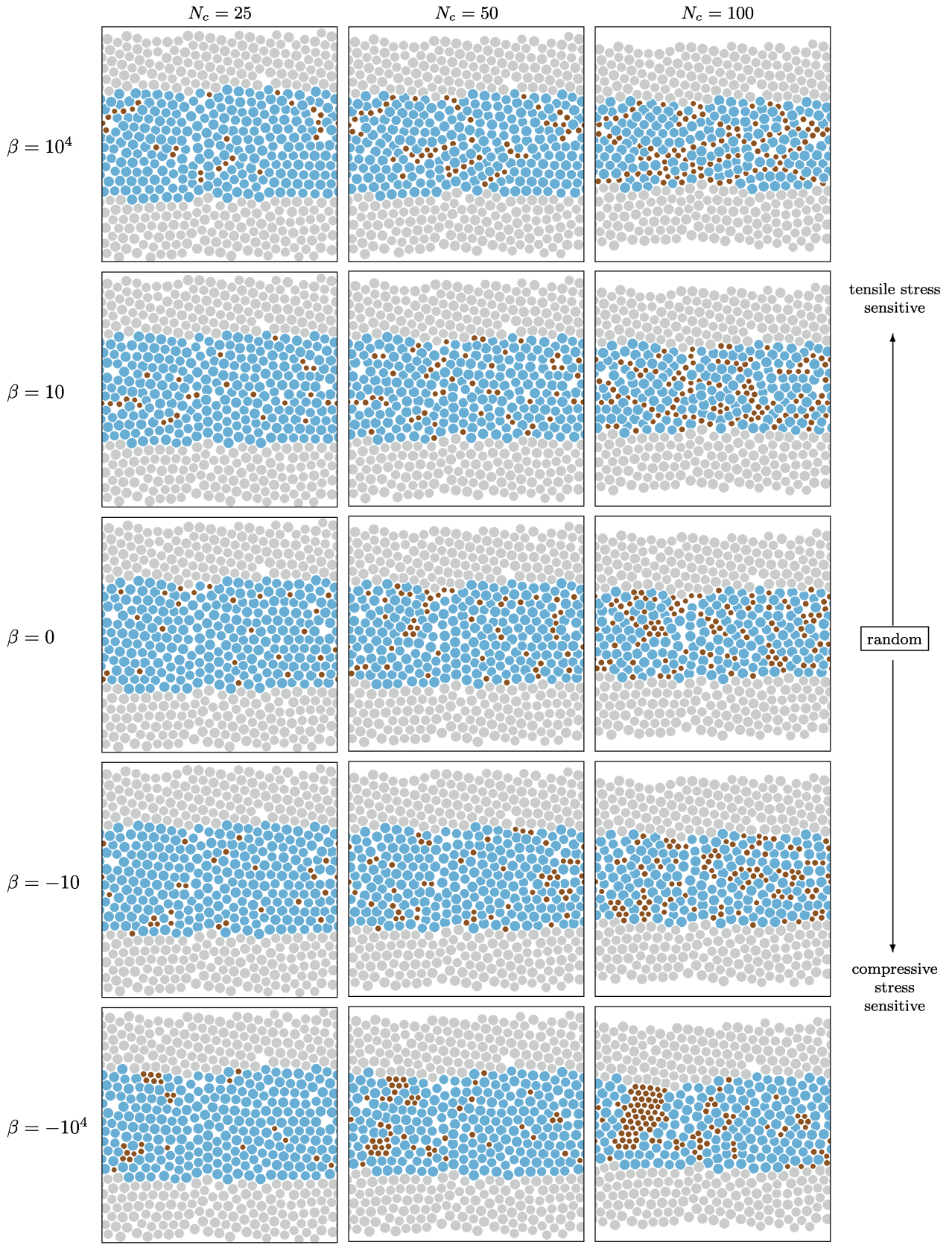}
  \caption{Comparison of the progression of cellular constrictions for
    different values of sensitivity stress parameter $\beta$ (as
    labeled) for $\shrinkIndex = 25$, $50$, and $100$.  Particle color
    key: active ($\aactive$) blue, inactive ($\inactive$) gray, and
    constricted ($\constricted$) brown.}
  \label{t12pbcx-betacomp}
\end{figure*}

\section{Active granular fluid model}
\label{Active granular fluid model}

\subsection{A simplified representation of an active cell
  monolayer}

The cellular constrictions that motivate our AGF model occur on the
outer surface of the embryonic cellular monolayer.  The interior (basal)
cellular ends remain relatively inactive throughout the initial slower
phase of apical constrictions.  Thus, to investigate coordination of
the constrictions via stress distribution we can use a simplified
description in which only the relevant ventral portion of the outer
surface of the embryo is explicitly represented.

We approximate this area of interest (i.e., the mesoderm primordium
and its immediate surroundings) as a two-dimensional plane; apical
cell ends are modeled as interacting active discs that constrict in a
stress-sensitive stochastic process.  In the past, complex systems of
strongly coupled particles (e.g., emulsions and foams) were
successfully modeled using interacting disks (or spheres)
\cite{Tewari-Schiemann-Durian-Knobler-Langer-Liu:1999,%
Durian:1995,%
Durian:1997,%
Langer-Liu:1997}.
We thus expect that using a closely packed
system of active disks to approximate the mechanically excitable cell
layer will reproduce the key features of the stress-driven
constriction process, and will yield valuable insights into
coordination of cellular constrictions by stress.

\subsection{System geometry}

Our system begins as a mechanically stable packing of $N$ discs
interacting via finite-range repulsive forces (which represent elastic
cell interactions).  The system, prepared using the algorithm
described in Appendix \ref{Preparation of initial condition}, occupies
a square simulation box of size $L$.  For a given particle number $N$
and disk diameters $\diameter_i$, the box size is determined from the
condition that the configuration is closely packed (area packing
fraction  $\phi\approx0.84$) and mechanically stable.

After the initial packing is prepared, we generate a list
$\neighborList$ of interacting neighbors.  We then add attractive
forces (representing cell adhesion) between the neighboring particles.
Particles $i$ and $j$ are assumed to be the interacting neighbors,
$(i,j)\in\neighborList$, if the condition
\begin{equation}
  \label{neighbor condition}
\rini_{ij}\le 1.1\diameter_{ij}
\end{equation}
is satisfied in the initial closely packed state, where $\rini_{ij}$
is the initial distance between the particles $i$ and $j$, and
$\diameter_{ij}=\frac{1}{2}(\diameter_i+\diameter_j)$ is their average
diameter.

In the initial state (i.e., before the disk constrictions occur), the
system is a disordered 50\% mixture of particles with the diameter
ratio $r=1.1$.  We use the bidisperse disk system to mimic
polydispersity of \Drosophila\ cells and to prevent formation of
hexagonal ordered structures in the initial closely packed state (such
structures are not observed for cells).  Subsequently, the system
undergoes a sequence of particle constrictions,
$\diameter_i\to\constrictionFactor\diameter_i$, according to the
algorithm described in Sec.\ \ref{Particle constriction protocol}.  In
our simulations we use the constriction factor
$\constrictionFactor=0.6$, corresponding to the size of constricted
cells \cite{Sweeton-Parks-Costa-Wieschaus:1991}.

\subsection{Active, inactive, and constricted particles}

The disks (see Fig.\ \ref{t12pbcx-betacomp}) are divided into three
distinct categories: active $\aactive$ (blue particles), inactive
$\inactive$ (gray), and particles already constricted $\constricted$
(brown).  Inactive discs cannot undergo constriction and will remain
the same size throughout the simulation.  Each active disk can
instantaneously constrict, and will do so by following the triggering
conditions described in Sec.\ \ref{Particle constriction protocol}
(the constricted disks cease to be active).

Particles that in the initial state are in the domain $0.25L<y<0.75L$
are active, and the remaining particles (in the regions $0<y<0.25L$
and $0.75L<y<L$) are inactive.  Our numerical simulations have been
performed for $N=512$ particles, so the initial stripe of the active
particles is approximately 11 particles wide, similar to the width of
the ventral region of active cells in a \Drosophila\ embryo
\cite{Sweeton-Parks-Costa-Wieschaus:1991}.

Our calculations are performed with periodic boundary conditions in
the horizontal (anteroposterior) direction $x$ and a free boundary
in the vertical (dorsoventral) direction $y$.  The use of the free boundary
condition is motivated by the relationship between the mesoderm
primordium (the black stripe in Fig. \ref{schematic}) and the cells on
the sides (the unmarked lateral region).  Cells of the lateral regions
are believed to passively deform and provide very little resistance to
the apical constrictions that occur in the mesoderm primordium
\cite{Sweeton-Parks-Costa-Wieschaus:1991,%
Polyakov-He-Swan-Shaevitz-Kaschube-Wieschaus:2014};
this condition is approximated by the free boundary condition.

\subsection{Interparticle potentials}

All particles interact via the finite-range, pairwise additive, purely
repulsive spring potential
\begin{equation}
  \label{repulsive potential}
  \Vrepulsion({r_{ij}}) = \frac{\epsilon}{2}(1 -
  {r_{ij}}/{\diameter _{ij}})^2\Theta ({\diameter _{ij}}/{r_{ij}} - 1),
\end{equation}
where $\epsilon$ is the characteristic energy scale, $r_{ij}$ is the
separation between particles $i$ and $j$, and $\Theta(x)$ is the
Heaviside step function.  In addition to the repulsion
\eqref{repulsive potential}, the neighboring particles
$i,j\in\neighborList$ interact via the attractive spring potential
\begin{equation}
  \label{attractive potential}
  \Vattraction({r_{ij}}) = \frac{\epsilon}{2}(1 -
  {r_{ij}}/{\diameter _{ij}})^2\Theta ({r_{ij}}/{\diameter_{ij}} - 1)
\end{equation}
that mimics the adhesion of neighboring cells.

In the initial state and after each particle constriction step, the
system is fully equilibrated (see Appendix \ref{System equilibration}
for details of the equilibration algorithm).  Thus, the sequence of
particle constriction steps is quasistatic.

\subsection{Evaluation of particle stress}

To characterize the overall stress exerted on particle $i$ by the
surrounding particles, we choose the following expression,
\begin{equation}
  \label{particle stress}
  \stress(i)=-\sum_{j\not=i}(\diameter_{ij}/\epsilon)\interparticleForce_{ij},
\end{equation}
where
\begin{equation}
  \label{central forces}
\interparticleForce_{ij}=-\diff \Vtot_{ij}/\diff r_{ij},
\end{equation}
with 
\begin{equation}
  \label{repulsive-attractive interparticle potential}
  \Vtot_{ij}(r_{ij})=\left\{
  \begin{array}{ll}
    \Vrepulsion(r_{ij})+\Vattraction(r_{ij}),&\quad (i,j)\in\neighborList\\
    \Vrepulsion(r_{ij}),&\quad (i,j)\not\in\neighborList
  \end{array}
  \right.
\end{equation}
are the interparticle central forces (which can be tensile,
$\interparticleForce_{ij}<0$, or compressive,
$\interparticleForce_{ij}>0$).  Accordingly, for $\stress(i)>0$ the
dimensionless particle stress \eqref{particle stress} is predominantly
tensile, and for $\stress(i)<0$ it is predominantly compressive.  The
particle stress \eqref{particle stress} is always evaluated for a
system in full mechanical equilibrium, in which there is no particle
motion, and all interparticle forces balance.

\subsection{Particle constriction protocol}
\label{Particle constriction protocol}

Particle constrictions are performed by iteratively repeating the
following procedure:
\begin{enumerate}
  \renewcommand{\theenumi}{\textit{\alph{enumi}}}
\item the system is fully equilibrated; 
  
\item particle stress $\stress(i)$ is evaluated according to
  Eq.\ \eqref{particle stress} for each particle $i$;

\item\label{constriction step} constriction probability (per one step)
  $P(i)$ is evaluated for each active (unconstricted)
  particle, $i\in\aactive$, according to Eq.\ \eqref{contraction
    probability with feedback} provided below;

\item diameters of the active  particles are decreased by
  a factor $\constrictionFactor$ with the probability
  $P(i)$.
\end{enumerate}
The above particle constriction step is repeated until the number of
constricted particles $\shrinkIndex$ reaches a 
prescribed limit.

\paragraph*{Constriction probabilities}

The probability $P(i,\shrinkIndex)$ of constriction of particle $i$ in
a system with $\shrinkIndex$ particles already constricted is
evaluated in step \ref{constriction step} of the above protocol from
the expression 
\begin{equation}
  \label{contraction probability with feedback}
  P(i,\shrinkIndex)
  =\frac{\alpha\left\{1
    +\beta\left[\stress(i,\shrinkIndex)/\stressMaxActive\right]^p\right\}}
     {(1+|\beta|)\activeIndex},
\end{equation}
where $\stress(i,\shrinkIndex)$ is the current value of the particle
stress \eqref{particle stress}, $\beta$ is the stress sensitivity
parameter, $\activeIndex$ is the current number of active particles,
and
\begin{equation}
  \label{maximal stress}
  \stressMaxActive=
  \max_{i\in\aactive}\left[\textrm{sign}(\beta)\stress(i,\shrinkIndex)\right]
\end{equation}
is the maximal tensile ($\beta>0$) or maximal compressive ($\beta<0$)
stress acting on these particles.  The parameter $\alpha>0$ sets the
average number of particles constricting in a single constriction
step, and the parameter $p>0$ (odd integer) determines the
sensitivity profile for the dependence of the constriction process on
the particle stress.  We use $\alpha=1$ and $p=3$ in all our
simulations.  The chosen non-unity value of $p$ gives a higher weight to the largest
(tensile or compressive) stresses.

The parameter $\beta$ determines the sign and magnitude of the overall
sensitivity of particle constrictions to the particle stress
$\stress(i)$.  We distinguish three fundamental cases:
\begin{enumerate}
  \newcommand{\itemc}[2]{\item\textit{#1: }{#2}}
  \itemc{uncorrelated random constrictions}{$\beta=0$;}

  \itemc{constrictions promoted by tensile stresses}{$\beta>0$;}

  \itemc{constrictions promoted by compressive stresses}{$\beta<0$.}
\end{enumerate}
In case 1, the system does not have any sensitivity to particle
stresses.  This purely random constriction process provides a
reference system for qualitatively describing how mechanical
sensitivity influences the propagation of constrictions through our
medium.  The other two triggering conditions introduce the sensitivity
of the constriction probability to tensile stresses (case 2) or
compressive stress (case 3).

The simulations for the stress-dependent cases were performed at a
variety of $\beta$ values.  Representative results, for $\beta=\pm 10$
(moderate stress sensitivity) and $\beta=\pm 10^4$ (strong stress
sensitivity), are discussed below.

\newcommand{\clearp}{\clearpage}
\renewcommand{\clearp}{}
\clearp
\begin{figure}
  \centering
  \includegraphics{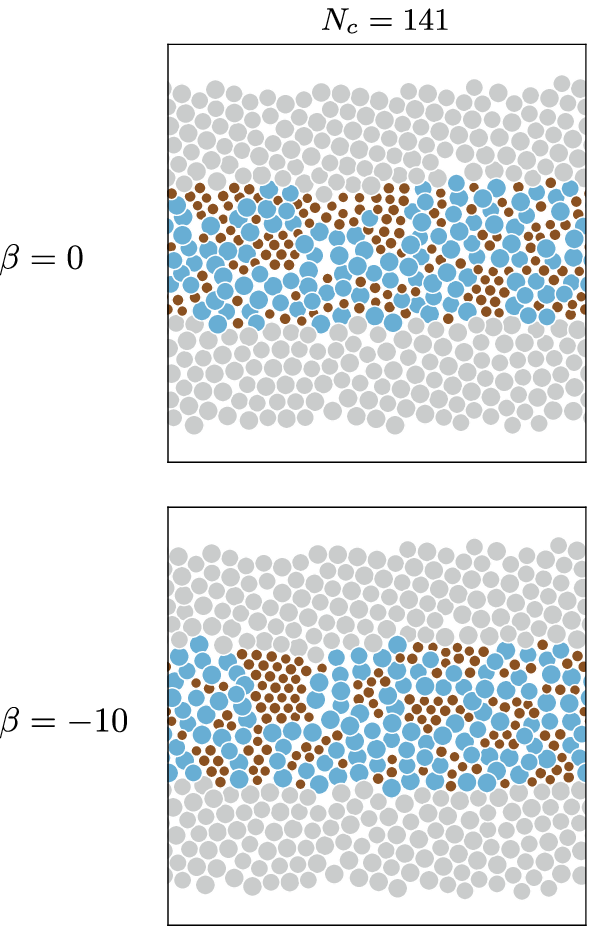}
  \caption{Same as Fig.\ \ref{t12pbcx-betacomp}, but for a larger value of
    $\shrinkIndex$.}
  \label{t2-betacomp_141}
\end{figure}

\clearp

\section{Results and Discussion}

Here we describe the results of our numerical simulations of
correlated particle constrictions in the AGF model introduced in
Sec.\ \ref{Active granular fluid model}.  We discuss the evolving
microstructure of constricted-particle regions and analyze the stress
distribution for different signs and magnitudes of the
stress sensitivity parameter $\beta$ in the constriction probability
function \eqref{contraction probability with feedback}, to determine
mechanisms that control the constriction patterns.

We focus on the initial stage of the system evolution, until
approximately 40\% of particles have constricted, because this stage
is most relevant to the initial phase of ventral furrow formation in
the \Drosophila\ embryo.  However, since our results may be relevant
also to other systems of mechanically active cells, we examine a
variety of constriction triggering conditions and provide a limited
set of results for longer times.

All simulations are performed for the same initial condition with
$\activeIndexInitial=258$ initially active particles.  Thus, for
$\shrinkIndex=25$ approximately 10\% particles have constricted, for
$\shrinkIndex=50$ approximately 20\%, and for $\shrinkIndex=100$
approximately 40\%.  

\subsection{Microstructural evolution}
\label{Microstructural evolution}

The results of our numerical simulations of the microstructural
evolution are summarized in Figs.\ \ref{t12pbcx-betacomp} and
\ref{t2-betacomp_141} [also see Movies 1(a)--1(e) in Supplementary Data].
The presented images of particle configurations reveal that the
spatial arrangement of the constricted particles shows a striking
dependence on the sign and magnitude of the constriction-triggering
stress.

In systems in which particle constrictions are induced by
\textit{tensile stress} (the top two rows of images in
Fig.\ \ref{t12pbcx-betacomp}), constricted particles form strongly
correlated \textit{chain-like structures}, which closely resemble CCCs
that we have identified in the \Drosophila\ embryo (see
Fig.\ \ref{cellconstchains}).  In contrast, the microstructure of the
system in which active particles are sensitive to \textit{compressive
  stresses} (the bottom two rows of images in
Fig.\ \ref{t12pbcx-betacomp}) is dominated by compact
  constricted-particle \textit{clusters}.

The chains that form for $\beta>0$ are partially aligned with the
$x$-direction (i.e., the direction of the system periodicity).  The
chains are initially disconnected (see the images for
$\shrinkIndex=25$ and 50), but at later times ($\shrinkIndex=100$ for
$\beta=10^4$) they grow into a percolating network
spanning the system in the $x$-direction.  This behavior closely
resembles the constricted-cell dynamics in the mesoderm primordium
shown in Fig.\ \ref{cellconstchains}.  Our results thus provide 
powerful (though indirect) evidence of the tensile-stress feedback
involved in cellular constrictions in the early phase of ventral
furrow formation in the \Drosophila\ embryo.

The chaining is most pronounced for large magnitudes of the stress
sensitivity parameter $\beta$.  We find that for a moderate value
$\beta=10$, the chains are fragmented and less aligned with the
$x$-axis than for $\beta=10^4$; and percolation occurs later, at
$\shrinkIndex\approx130$ [see Movie 1(b) in
Supplementary Data].

In systems with particle constrictions induced by compressive
stresses, $\beta<0$, we observe the formation of compact clusters,
with significant size polydispersity and inhomogeneous spatial
distribution.  The clusters remain disconnected and do not percolate,
until much later in the process.

For a moderate sensitivity-parameter value $\beta=-10$, the clusters
are much smaller than for the system with strong stress sensitivity
$\beta=-10^4$.  In fact, during the evolution stage depicted in
Fig.\ \ref{t12pbcx-betacomp}, the cluster distribution in systems with
$\beta=-10$ and $\beta=0$ (no stress sensitivity) looks similar;
however, at  later stages the difference between the fully random and
weak-stress-sensitivity systems becomes much larger (see
Fig.~\ref{t2-betacomp_141}).

We note that even in a fully random case, $\beta=0$, a significant
number of clusters form already at a relatively early stage of evolution
(see Fig.\ \ref{t12pbcx-betacomp} for $\shrinkIndex=50$).  This is
because an isolated constricted particle typically has several active
neighbors, which increases the probability of formation of small
groups of constricted particles in an uncorrelated random process.

\begin{figure}
  \centering 
  \includegraphics{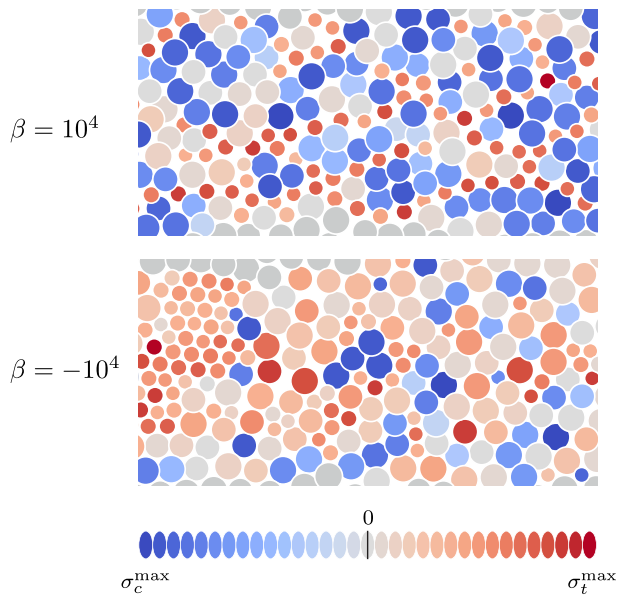}
  \caption{A visual representation of the distribution of stress
    $\stress$ through the active region of a granular fluid at
    $\shrinkIndex=100$ and $\beta$ as labeled.  Shades of blue and red
    show a net compressive and tensile stress, respectively.  (The
    color scale is relative to the maximal compressive and tensile
    stresses in the system $\sigma^\maxx_\compressive$ and
    $\sigma^\maxx_\tensile$.)}
  \label{t12pbcx-stressdist}
\end{figure}

\clearp
\begin{figure*}[b]
   \centering
   \includegraphics{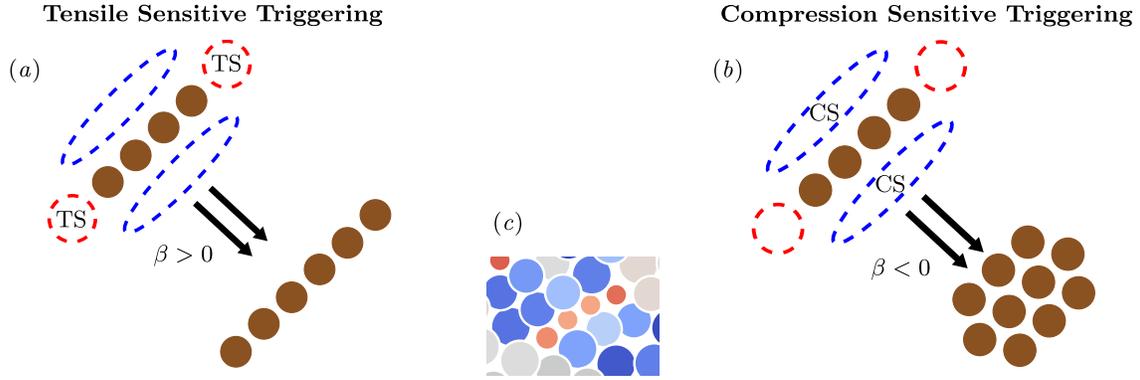}
   \caption{Mechanism of formation of \subfig{a} chains and
     \subfig{b} clusters of constricted particles (brown circles);
     and \subfig{c} the distribution of particle stress (color scale
     as in Fig.\ \ref{t12pbcx-stressdist}) near a chain of four
     constricted particles. Areas of tensile stress (TS, red dashed
     circles) and areas of compressive stress (CS, blue dashed
     ovals) denote regions of tensile-stress- and
     compressive-stress-triggered constrictions, respectively.}
   \label{microstructure evolution}
  
\end{figure*}

\clearp
\begin{figure*}
  \centering 
  \includegraphics{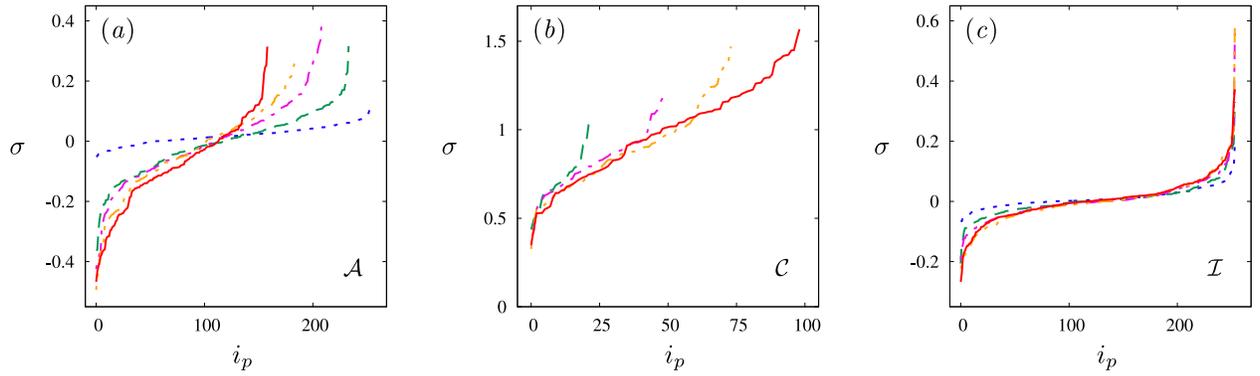}
  \caption{Stress distribution $\stress=\stress(i,\shrinkIndex)$ for
    \subfig{a} active, \subfig{b} constricted, and \subfig{c} inactive
    particles, in a system sensitive to tensile stress ($\beta=10^4$),
    with $\shrinkIndex=0$ (blue dotted line), $\shrinkIndex=25$ (green
    dashed), $\shrinkIndex=50$ (purple dash--dot), $\shrinkIndex=75$
    (orange dash--dot--dot), and $\shrinkIndex=100$ (red solid)
    constricted particles.  For a given $\shrinkIndex$, the results are
    sorted by the increasing value of stress and shown vs.\ particle index
    $\particleIndex$.}
  \label{stress distribution type 1x}
\end{figure*}

\clearp
\begin{figure*}
  \centering 
  \includegraphics{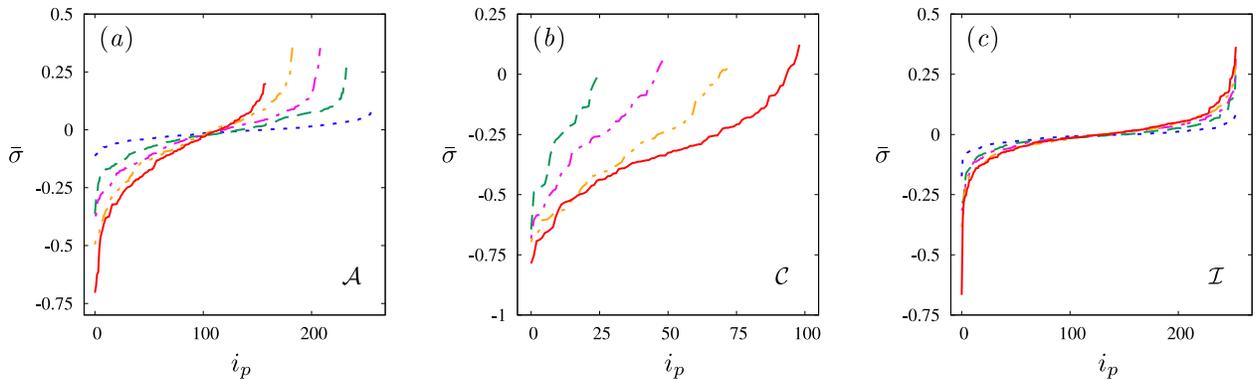}
  \caption{Same as Fig.\ \ref{stress distribution type 1x}, but for a
    system sensitive to compressive stress ($\beta=-10^4$), and for
  the values of the particle pressure $\pressure=-\stress$.}
  \label{stress distribution type 2x}
\end{figure*}

\clearp
\begin{figure}
  \centering
  \includegraphics{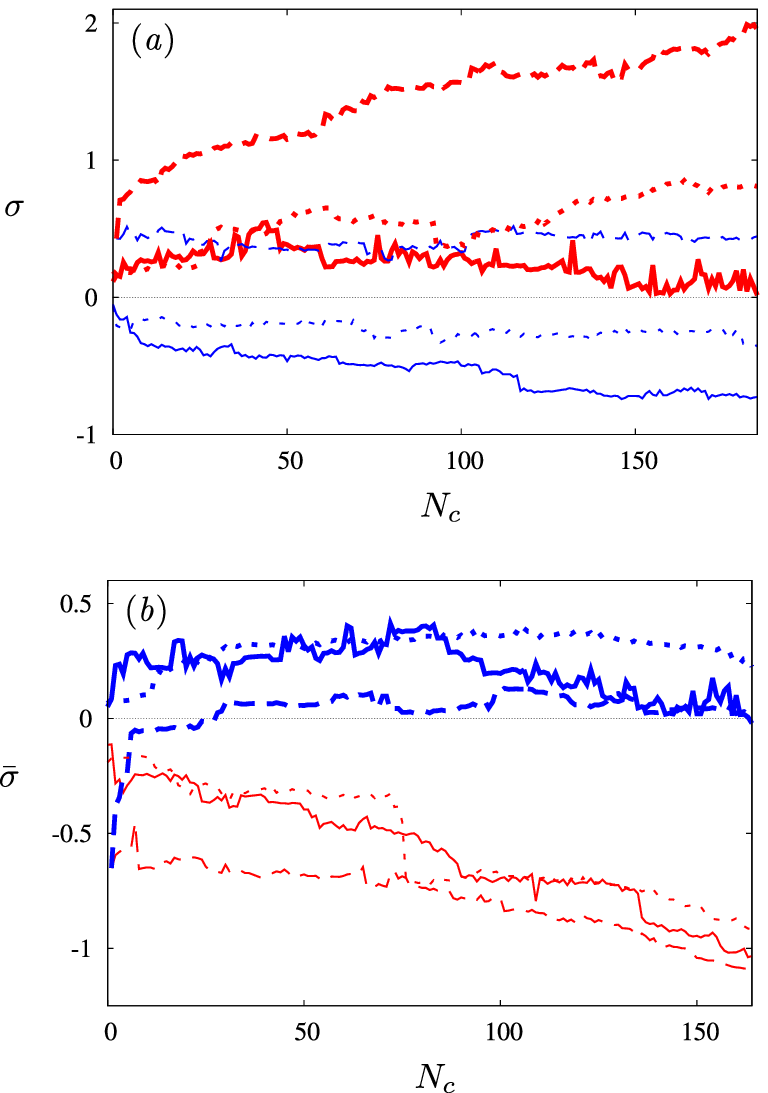}
  \caption{\subfig{a} Evolution of the maximal  stress
    $\stress=\stressMax$ (thick lines) and minimal  stress
    $\stress=\stressMin$ (thin lines) in a system with stress
    sensitivity parameter $\beta=10^4$ for the three types of
    particles: active (solid lines), constricted (dashed), and
    inactive (dotted).  \subfig{b} The same as \subfig{a}, except that
    for a system with $\beta=-10^4$ and for the values of particle
    pressure $\pressure=-\stress$.  In both panels, the red lines
    correspond to the tensile end of the stress range, and blue ones
    to the compressive end.}
  \label{stress evolution type pbcx}
\end{figure}

\clearp

\subsection{The mechanism of chain and cluster formation}

To elucidate the mechanisms of chain and cluster formation, we examine
the stress distribution in the evolving AGF undergoing particle
constrictions.  A color map of the particle stress,
Eq.\ \eqref{particle stress}, is presented in
Fig.\ \ref{t12pbcx-stressdist} for a system with constrictions
triggered by tensile stresses (top panel, $\beta=10^4$) and
compressive stresses (bottom panel, $\beta=-10^4$).  Both panels show
the same stage of the evolution, $\shrinkIndex=100$.  The images were
obtained by recoloring the corresponding panels in
Fig.\ \ref{t12pbcx-betacomp}, to visualize the distribution of stress
in the active particle band.  Movies 2(a) and 2(b) in Supplementary
Data show the evolution of the stress distribution during the
constriction process.

A close examination of the top panel of Fig.\ \ref{t12pbcx-stressdist}
reveals the following key features of the stress distribution that are
essential for understanding the microstructural evolution (see the
schematics in Fig.\ \ref{microstructure evolution}):
\begin{enumerate}
  \item chains of constricted particles are predominantly subject to
    tensile stress;
  \item unconstricted particles that are positioned alongside the
    chain are predominantly compressed, while those near the chain
    ends are predominantly under tension. 
\end{enumerate}

In a system sensitive to tensile stresses (Fig.\ \ref{microstructure
  evolution}\subfig{a}), active particles are most likely to constrict
near the ends of an already formed chain, because the tensile stress
is predominant in these regions (as indicated by the dashed red
circles in Fig.\ \ref{microstructure evolution}\subfig{a} and the
tensile stress coloring in Fig.\ \ref{microstructure
  evolution}\subfig{c}).  Thus the chain increases in length.  This
mechanism not only promotes chain growth, but also results in
increased chain connectivity, stimulating the expansion of a
constricted-particle-chain network, and leading to its eventual
percolation.

In a system sensitive to compressive stresses (see
Fig.\ \ref{microstructure evolution}\subfig{b}), the above mechanism
of growth of a constricted region is inverted: the compressed
particles alongside the chain (i.e., in the regions indicated by the
dashed blue ovals) are now likely to constrict, which results in
restructuring of the chain into a compact cluster.  A similar mechanism
governs growth of already formed clusters.

\subsection{Evolution of the particle stress distribution}
\label{Stress distribution}

A comparison of the stress distribution for a tensile-stress-sensitive
system (top panel of Fig.\ \ref{t12pbcx-stressdist}) and for a
compressive-stress-sensitive system (bottom panel of
Fig.\ \ref{t12pbcx-stressdist}) shows that the stresses are
distributed very differently.  For tension-sensitive triggering, the
tensile stress propagates along a network of constricted-particle
chains, leaving pockets of compressive stress between the chains.  In
contrast, for compression-sensitive triggering (i.e., when a network
of constricted particles does not form), the tensile stress
distribution is much more uniform, and, moreover, the tensile stresses
are predominant, and only small pockets of compressive stress remain.

To gain further insights into stress feedback mechanisms that may be
important in ventral furrow formation (and more generally in behavior
of active particulate systems), we examine the evolution of particle
stress for different particle populations.  The distribution of
particle stress for the populations of active particles $\aactive$,
constricted particles $\constricted$, and inactive particles
$\inactive$ is shown in Fig.\ \ref{stress distribution type 1x} for a
tensile-sensitive system with $\beta=10^4$.  Figure \ref{stress
  distribution type 2x} shows the corresponding results for the particle
pressure
\begin{equation}
  \label{particle pressure}
  \pressure=-\stress
\end{equation}
in a compression-sensitive system with $\beta=-10^4$.  (We call
quantity \eqref{particle pressure} the particle pressure, in line with
the standard convention in fluid mechanics, in which the pressure
tensor is the negative of the stress tensor.)

In both cases the results are sorted from the smallest to largest
value of the stress $\stress$ (pressure $\pressure$) and shown
vs.\ particle index $\particleIndex$. Accordingly, the largest
triggering stress (pressure) corresponds to the upper-right end of the
curves representing the stress (pressure) distribution.

The results reveal several important features common to all particle
populations and both constriction-triggering conditions.  First, the
slope of the curves sharply increases near the ends of the curves
$\stress(\particleIndex)$ and $\pressure(\particleIndex)$. This
behavior indicates that there exist small particle subsets for which
the stress $\stress$ (or pressure $\pressure$) significantly differs
from the stress (pressure) for typical particles.  Second, the stress
distribution is initially narrow, and becomes significantly wider when
particles start to constrict creating local inhomogeneities within the
medium, and generating large positive and negative values of particle
stress.  The latter feature of the stress distribution is also visible
in Fig.\ \ref{stress evolution type pbcx}, where the minimal and
maximal particle stress
\begin{subequations}
  \label{min-max stress and pressure}
\begin{equation}
  \label{min-max  stress}
  \stressMin(\shrinkIndex)=\min_{i\in\population}\stress(i,\shrinkIndex),\qquad
  \stressMax(\shrinkIndex)=\max_{i\in\population}\stress(i,\shrinkIndex),
\end{equation}
and particle pressure
\begin{equation}
  \label{minmax pressure}
  \pressureMin=-\stressMax,\qquad
  \pressureMax=-\stressMin
\end{equation}
\end{subequations}
in the population $\population=\aactive, \constricted, \inactive$ are
plotted versus the number of constricted particles for tension- and
compression-sensitive systems.

\paragraph*{Tensile-stress-sensitive triggering} The results depicted
in Figs.\ \ref{stress distribution type 1x}\subfig{a} and \ref{stress
  evolution type pbcx}\subfig{a} show that the maximal tensile stress
in the population of active particles $\aactive$ in a
tension-sensitive medium initially increases, but subsequently
gradually decreases, and eventually vanishes.  The non-monotonic
behavior of the tensile stress in the population $\aactive$ is a
consequence of the formation of a network of connected
constricted-particle chains which support most of the tensile stress
(see Fig.\ \ref{t12pbcx-betacomp} and the top panel of
Fig.\ \ref{t12pbcx-stressdist}).

Because the constriction probability \eqref{contraction probability
  with feedback} is normalized by the maximal stress \eqref{maximal
  stress} in the current configuration, our simulations cannot be
continued beyond the point at which the maximal tensile stress in the
active population $\aactive$ vanishes.  In a modified model with
normalization by a fixed characteristic stress, the constriction
process could be continued, but would significantly slow down at the
point $\stressMaxActive\approx0$.  (In  wild type \Drosophila\ the
slowdown would not occur because of the earlier transition to the
rapid phase of apical constrictions, which is controlled by different
mechanisms; the slowdown, however, could perhaps be observed in
\Drosophila\ mutants.)

The growth of the stress-supporting network of constricted particles
is reflected in a steadily increasing stress in the particle
population $\constricted$ (see Fig.\ \ref{stress distribution type
  1x}\subfig{b} and the dashed red line in \ref{stress evolution type
  pbcx}\subfig{a}).  We hypothesize that the stress-supporting network
of CCCs in the \Drosophila\ embryo (analogous to the interconnected
constricted-particle chains in our AGF model) plays a biologically
useful role.  First, since the chains distribute stresses non-locally
in the entire active region, they may mitigate the effect of decreased
cell contractility in some domains (such domains may result from
random fluctuations or genetic defects).  Second, a tightly stretched
band of interconnected CCCs may help to organize a coherent tissue
motion at the onset of the second phase of ventral furrow
formation. In both cases, CCCs would contribute to robustness of the
invagination process.  However, the role of CCCs requires further
investigations.

\paragraph*{Compressive-stress-sensitive triggering}

As discussed at the beginning of Sec.\ \ref{Stress distribution},
compression-sensitive systems develop large continuous areas of
tensile stress.  These areas include compact domains of active and
constricted particles.  As a consequence of this morphology, the
tensile stress (i.e., the negative pressure $\pressure$ in the plots
shown in Figs.\ \ref{stress distribution type 2x} and \ref{stress
  evolution type pbcx}\subfig{b}) is similarly distributed in the
populations $\aactive$ and $\constricted$.  As seen in
Fig.\ \ref{stress evolution type pbcx}\subfig{b}, the tensile stress
is somewhat larger for constricted particles (red dashed line) than for
active particles (red solid line), but the difference is much smaller
than the corresponding difference for tension-sensitive triggering
(see dashed and solid red lines in Fig.\ \ref{stress evolution type
  pbcx}\subfig{a}).

The compressive stress (i.e., the positive pressure $\pressure$) is
less evenly distributed.  In the population of active particles
$\aactive$, the maximal value $\pressureMax$ is relatively small and
decreases to zero at long times.  Only a few constricted particles are
under compression according to the stress map shown in the bottom
panel of Fig.\ \ref{t12pbcx-stressdist}, and the maximal pressure is
negative or close to zero during the entire evolution.

We note that the behavior of the stress distribution within the
population of constricted particles $\constricted$ is qualitatively
different in the compression-sensitive medium (see Fig.\ \ref{stress
  distribution type 2x}\subfig{b}) and the tension-sensitive medium
(see Fig.\ \ref{stress distribution type 1x}\subfig{b}).  In the
former case, the slope of the curve
$\pressure=\pressure(\particleIndex)$ decreases with the increasing
$\shrinkIndex$, and in the latter case the slope of the corresponding
curve $\stress=\stress(\particleIndex)$ remains constant (only the
middle, nearly linear, part of the curve varies in length).  This
universal slope is likely to be a signature of scaling properties of
the developing constricted-particle network.

\paragraph*{Behavior of the inactive region}  As depicted in
Figs.\ \ref{stress distribution type 1x}\subfig{c} and \ref{stress
  distribution type 2x}\subfig{c}, the stress distribution in the
inactive-particle region $\inactive$ is relatively featureless.  Most
of the particles experience small positive or negative stresses, and,
according to stress maps (not shown) only a small number of particles
on the border between the active and inactive regions are affected by
particle constrictions. The maximal tensile stress grows with
$\shrinkIndex$, and the compressive stress saturates (see
Fig.\ \ref{stress evolution type pbcx}).

With the free boundary condition in the $y$-direction, the inactive 
particles do not affect the active region in a significant way.
However, for more resistive boundary conditions (e.g., periodic
boundary conditions in both $x$ and $y$ directions; results not
shown), the interaction between the passive and active regions is much
stronger.  In our future studies, this effect will be investigated in
the context of interactions between active cells in the mesoderm
primordium and the surrounding cells in more lateral regions.

\section{Conclusions}

Mechanical stress fields are now believed to play a pivotal role in
many biological developmental processes.  Therefore, it is crucial to
establish methods to investigate local cell--cell mechanical
interactions and global stress distributions across tissues and
evaluate the effect of such local and global phenomena on mechanical
cell activity. We have shown that modeling a tissue as an active
granular medium can offer a means of analyzing mechanical feedback
involved in tissue development.

Our AGF model of apical constrictions in the \Drosophila\ embryo during
the early stage of ventral furrow formation has demonstrated
constriction patterns that are qualitatively similar to those observed
\textit{in vivo}.  The key new element of the  model is the
quantification of the mechanical sensitivity of cells to tensile
stresses, which are responsible for an increase in the cell
constriction probability.  The agreement between the model predictions
and constriction patterns observed in the \Drosophila\ mesoderm
primordium provides evidence of the role of mechanical feedback in
the early stage of morphogenesis examined here.

We have considered a wide range of constriction triggering conditions
and analyzed the associated stress distribution, which evolves as the
cellular constriction process progresses.  We have also shown that in
systems in which cells are sensitive to compressive stresses, growing
clusters of constricted particles are formed instead of
constricted-particle chains.  It follows that mechanical feedback can
be used to control the system morphology.

The cell dynamics during the initial apical constriction phase of
ventral furrow formation (considered here) can be modeled using a 2D
AGF approach, because cells at this stage are mechanically active only
in a narrow, nearly planar region.  However, subsequent morphogenetic
movements (e.g., the later phase of ventral furrow formation, cephalic
furrow formation, and germ band extension) involve large-scale
collective cell motions in different mechanically coupled domains.
Understanding the role of mechanical feedback in coordinating cell
activity will thus require development of comprehensive full-embryo 3D
models in which motion of all cells (approximately 6000), arranged in
an epithelial monolayer surrounding the yolk sac, will be explicitly
followed; we are working on such models.

We expect that a variety of morphogenetic and organogenetic processes
can be studied by modeling a developing tissue as an active granular
fluid.  For example, mechanical stresses have been shown to influence
multiple aspects of heart development in Zebrafish.  Stress exerted
upon cells by fluid flow influences the number of chambers which are
developed
\cite{Hove-Koster-Forouhar-AcevedoBolton-Fraser-Gharib:2003},
valve growth
\cite{Vermot-Forouhar-Liebling-Wu-Plummer-Gharib-Fraser:2009},
and the establishment of pacemaker cells.  It is possible that these
mechanically sensitive aspects of cardiogenesis can be evaluated by
considering an appropriate AGF model.

Statistical mechanics methods and simulation techniques that were initially
developed for investigations of complex fluids have significantly
contributed to the understanding of molecular-level mechanisms in
biological systems.  In particular, fundamental studies of protein
folding
\cite{Bryngelson-Wolynes:1987,%
Li-Scheraga:1987,%
Lau-Dill:1989}
(also advanced by George Stell's group
\cite{Foffi-McCullagh-Lawlor-Zaccarelli-Dawson-Sciortino-Tartaglia-Pini-Stell:2002,%
Lee-Stell-Wang:2003,%
Leite-Onuchic-Stell-Wang:2004}%
) have led to the development of designer proteins
\cite{Davis-Chin:2012,%
Mandell-Lajoie-Mee-Takeuchi-Kuznetsov-Norville-Gregg-Stoddard-Church:2015,%
Procko-Hedman-Hamilton-Seetharaman-Fleishman-Su-Aramini-Kornhaber-Hunt-Tong-Montelione-Baker:2013}.
Other examples of cross-pollination between fluid-state physics and biology include
rapid progress in areas such as cytoskeleton dynamics
\cite{Jain-Inamdar-Padinhateeri:2015,%
Heussinger-Bathe-Frey:2007,%
Mizuno-Tardin-Schmidt-MacKintosh:2007}
and behavior of cell membranes
\cite{Baumgart-Hammond-Sengupta-Hess-Holowka-Baird-Webb:2007,%
Baumgart-Hess-Webb:2003,%
Lingwood-Simons:2010}.

We anticipate that studies of the collective phenomena associated with
mechanical cellular activity and inter-cellular interactions, including
results of multicellular modeling of mechanical feedback during tissue
formation, will be of similar importance for understanding
morphogenesis and organogenesis.  Further, the knowledge gained
will also lead to applications in tissue engineering.

\begin{acknowledgments}
GJG gratefully acknowledges financial support from NTU startup funding
104R7417.  Imaging experiments were supported by funds from TTUHSC to
JHT.

\end{acknowledgments}

\appendix

\section{Imaging a \Drosophila\ embryo}
\label{experimental methods}

The ventral surfaces of live \textit{Drosophila melanogaster} embryos were
imaged to observe the constriction of cell apices during ventral
furrow formation
\cite{Martin-Kaschube-Wieschaus:2009}.
Cell apices were visualized by the fluorescently labeled plasma
membrane protein encoded by the \textit{Spider-GFP} transgene
\cite{Morin-Daneman-Zavortink-Chia:2001}.
Embryos were prepared for imaging, selected by age under the
dissecting microscope, oriented and glued to coverslips as described
\cite{Martin-Kaschube-Wieschaus:2009,%
  Cavey-Lecuit:2008,%
  Spencer-Siddiqui-Thomas:2015}.
To avoid any artifacts caused by gluing the vitelline membrane of the
ventral surface of the embryo to the coverslip that we imaged through,
we designed and constructed an imaging chamber. The bottom of
the chamber consisted of a coverslip with a strip of embryo glue
flanked on either side by two layers of double-sided Scotch tape. Two
layers of double-sided Scotch tape are sufficient to avoid compression
of the embryo
\cite{Figard-Sokac:2011}.
The dorsal sides of the embryos were glued to the bottom of the
imaging chamber and the embryos were covered with halocarbon oil 27
(Sigma). A number 1.5 coverslip was adhered to the double-sided tape
to close the chamber and allow the ventral surfaces of the embryos to
be visualized. The imaging chamber was taped to a glass slide. Images
were collected every 10 seconds using a Zeiss Axio Imager.A1
microscope with Axiovision 4.4 software. Time-lapse images were
compiled using Axiovision and ImageJ
\cite{Schneider-Rasband-Eliceiri:2012}.
Images were processed and constricted apices marked using Photoshop.
Constricted apices were indicated based on the widths in the smallest
dimension of the cell apex and its evolution over time.

\section{Numerical simulation details}
\label{Numerical-simulation details}

\subsection{Preparation of the initial disk configuration}
\label{Preparation of initial condition}

To prepare the initial disk configuration for our numerical
simulations of the correlated apical constriction process, we
\subfig{a} generate a random close packing (RCP) of frictionless disks
interacting via the finite-range repulsive potential \eqref{repulsive
  potential}; \subfig{b} establish the neighbor list $\neighborList$
according to the criterion \eqref{neighbor condition}; \subfig{c} add
the attractive potential \eqref{attractive potential}; and \subfig{d}
equilibrate the system.

The required RCP of disks is prepared by following the
packing-generation procedure described in \cite{gao06}.  Accordingly,
the disks are randomly placed in a square unit cell with periodic
boundary conditions. The particle diameters are then increased or
decreased by a gradually decreasing factor, to remove overlaps or gaps
between particles; the particle size change is followed by energy
minimization.  The process is repeated until there is no room to
change the size of particles without creating an overlap \cite{gao06}.

\subsection{System equilibration}
\label{System equilibration}

Nonequilibrium configurations arising during the initial-state
generation and after each particle contraction are equilibrated
using molecular dynamics of a dissipative system with the 
interparticle central potential forces
\begin{equation}
  \label{interparticle force vector}
\InterparticleForce_{ij}=\interparticleForce_{ij}\RadialUnitVector_{ij},
\end{equation}
and velocity-dependent dissipative resistance
forces $\ResistanceForce_{ij}$.  Accordingly, the equations of motion
\begin{equation}
  \label{eqn_of_motion}
  m_i\mathbf{a}_i
  =\sum_{j\not=i}(\InterparticleForce_{ij}+\ResistanceForce_{ij})
\end{equation}
(where $m_i$ and $\mathbf{a}_i$ are the mass and acceleration of
particle $i$, and $\RadialUnitVector_{ij}$ is the unit vector pointing
from the center of particle $j$ to $i$) are solved using the velocity
Verlet algorithm, until the system reaches the energy minimum.

During the initial packing preparation we use
\begin{equation}
  \label{only repulsive force}
\interparticleForce_{ij}=-\diff \Vrepulsion({r_{ij}})/\diff r_{ij}
  \end{equation}
and
\begin{equation}
  \label{friction for overlap}
  \ResistanceForce_{ij}=
  -b\Theta ({\diameter_{ij}}/{r_{ij}} - 1)
  (\bv_{ij}\bcdot\RadialUnitVector_{ij})\RadialUnitVector_{ij},
\end{equation}
where $\bv_{ij}=\bv_i-\bv_j$ is the relative velocity between
particles $i$ and $j$, and $b$ is the resistance constant.  During the
subsequent process of particle constrictions we use the potential
force \eqref{central forces} and
\begin{equation}
  \label{resistance with neighbors included}
  \ResistanceForce_{ij}=\left\{
  \begin{array}{ll}
    -b(\bv_{ij}\bcdot\RadialUnitVector_{ij})\RadialUnitVector_{ij},
    &\quad (i,j)\in\neighborList\\
    -b\Theta ({\diameter_{ij}}/{r_{ij}} - 1)
    (\bv_{ij}\bcdot\RadialUnitVector_{ij})\RadialUnitVector_{ij}
    &\quad (i,j)\not\in\neighborList
  \end{array}
  \right.
  \end{equation}
(i.e., attractive and dissipative interactions between non-overlapping
neighbors are also included).

The equilibration is performed with $b=0.5$ in the dimensionless units
where the diameter $\diameterSmall$ and mass $\massSmall$ of the small
particles and the interparticle potential amplitude $\epsilon$ are
chosen as the reference length, mass, and energy scales.  The mass of
the large particles is $\massLarge=\massSmall r^2$, where $r$ is the
initial diameter ratio; the particle masses $\massSmall$ and $\massLarge$
are not affected by constrictions.  We note that the particle masses
and the specific form of the dissipative resistance forces influence
only the numerical efficiency of the equilibration process, but do
not affect the final equilibrated state.

\renewcommand{\wherebib}{BIB}
\bibliographystyle{unsrt}
 \bibliography{\wherebib/drosophila,%
   \wherebib/genbiology,%
   \wherebib/granularmedia,%
   BIB/numerical}

\providecommand{\noopsort}[1]{}\providecommand{\singleletter}[1]{#1}%
\begin{thebibliography}{10}

\bibitem{Mammoto-Ingber:2010}
T.~Mammoto and D.E. Ingber.
\newblock {Mechanical control of tissue and organ development}.
\newblock {\em Development}, 137(9):1407--1420, 2010.

\bibitem{Miller-Davidson:2013}
C.J. Miller and L.A. Davidson.
\newblock {The interplay between cell signalling and mechanics in developmental
  processes}.
\newblock {\em Nat. Rev. Genet.}, 14(10):733--744, 2013.

\bibitem{Zhang-Labouesse:2012}
H.~Zhang and M.~Labouesse.
\newblock {Signalling through mechanical inputs - a coordinated process}.
\newblock {\em J. Cell Sci.}, 125(13):3039--3049, 2012.

\bibitem{Shiu-Weiss-Hoying-Iwamoto-Joung-Quam:2005}
Y.-T. Shiu, J.A. Weiss, J.B. Hoying, M.N. Iwamoto, I.S. Joung, and C.T. Quam.
\newblock The role of mechanical stresses in angiogenesis.
\newblock {\em Crit. Rev. Biomed. Eng.}, 33(5):431--510, 2005.

\bibitem{Farge:2003}
E.~Farge.
\newblock {Mechanical induction of twist in the Drosophila foregut/stomodeal
  primordium}.
\newblock {\em Curr. Biol.}, 13(16):1365--1377, 2003.

\bibitem{Pouille-Ahmadi-Brunet-Farge:2009}
P.-A. Pouille, P.~Ahmadi, A.-C. Brunet, and E.~Farge.
\newblock {Mechanical Signals Trigger Myosin II Redistribution and Mesoderm
  Invagination in Drosophila Embryos}.
\newblock {\em Sci. Signal.}, 2(66), 2009.

\bibitem{Idema-Liu:2013}
T.~Idema and A.J. Liu.
\newblock Mechanical signaling via nonlinear wavefront propagation in a
  mechanically-excitable medium.
\newblock {\em Phys. Rev. E}, 89:062709, 2014.

\bibitem{Idema-Dubuis-Kang-Manning-Nelson-Lubensky-Liu:2013}
T.~Idema, J.O. Dubuis, L.~Kang, M.L. Manning, P.C. Nelson, T.C. Lubensky, and
  A.J. Liu.
\newblock The syncytial drosophila embryo as a mechanically excitable medium.
\newblock {\em PLoS ONE}, 8(10):e77216, 2013.

\bibitem{Sweeton-Parks-Costa-Wieschaus:1991}
D.~Sweeton, S.~Parks, M.~Costa, and E.~Wieschaus.
\newblock Gastrulation in \textit{Drosophila}: the formation of the ventral
  furrow and posterior midgut invaginations.
\newblock {\em Development}, 112:775--789, 1991.

\bibitem{Howell-Behringer-Veje:1999}
D.W. Howell, R.P. Behringer, and C.T. Veje.
\newblock {Fluctuations in granular media}.
\newblock {\em Chaos}, 9(3):559--572, 1999.

\bibitem{Behringer-Howell-Kondic-Tennakoon-Veje:1999}
R.P. Behringer, D.~Howell, L.~Kondic, S.~Tennakoon, and C.~Veje.
\newblock {Predictability and granular materials}.
\newblock {\em Physica D}, 133(1-4):1--17, 1999.

\bibitem{Xie-Martin:2015}
S.~Xie and A.C. Martin.
\newblock {Intracellular signalling and intercellular coupling coordinate
  heterogeneous contractile events to facilitate tissue folding}.
\newblock {\em Nat. Commun.}, 6:7161, 2015.

\bibitem{Leptin:1999}
M.~Leptin.
\newblock Gastrulation in drosophila: the logic and the cellular mechanisms.
\newblock {\em EMBO J.}, 18(10):3187--3192, 1999.

\bibitem{StJohnston-NussleinVolhard:1992}
D.~St~Johnston and C.~N{\"u}sslein-Volhard.
\newblock The origin of pattern and polarity in the drosophila embryo.
\newblock {\em Cell}, 68:201--219, 1992.

\bibitem{Riechmann-Ephrussi:2001}
V.~Riechmann and A.~Ephrussi.
\newblock Axis formation during drosophila oogenesis.
\newblock {\em Curr. Opin. Genet. Dev.}, 11:374--383, 2001.

\bibitem{Huynh-StJohnston:2004}
J.-R. Huynh and D.~St~Johnston.
\newblock The origin of asymmetry: Early polarisation of drosophila germline
  cyst and oocyte.
\newblock {\em Curr. Biol.}, 14:R438--R449, 2004.

\bibitem{Vaneeden-StJohnston:1999}
F.~van Eeden and D.~St~Johnston.
\newblock The polarisation of the anterior-posterior and dorsal-ventral axes
  during drosophila oogenesis.
\newblock {\em Curr. Opin. Genet. Dev.}, 9:396--404, 1999.

\bibitem{Leptin-Grunewald:1990}
M.~Leptin and B.~Grunewald.
\newblock {Cell-Shape Changes During Gastrulation in Drosophila}.
\newblock {\em Development}, 110(1):73--84, 1990.

\bibitem{Leptin:1991}
M~Leptin.
\newblock {twist and snail as positive and negative regulators during
  Drosophila mesoderm development}.
\newblock {\em Genes \& Development}, 5(9):1568--1576, 1991.

\bibitem{Ip-Maggert-Levine:1994}
Y.T. Ip, K.~Maggert, and M.~Levine.
\newblock {Uncoupling gastrulation and mesoderm differentiation in the
  Drosophila embryo}.
\newblock {\em EMBO J.}, 13(24):5826--5834, 1994.

\bibitem{Seher-Narasimha-Vogelsang-Leptin:2007}
T.C. Seher, M.~Narasimha, E.~Vogelsang, and M.~Leptin.
\newblock {Analysis and reconstitution of the genetic cascade controlling early
  mesoderm morphogenesis in the Drosophila embryo}.
\newblock {\em Mechanisms of Development}, 124(3):167--179, 2007.

\bibitem{Martin-Kaschube-Wieschaus:2009}
A.C. Martin, M.~Kaschube, and E.F. Wieschaus.
\newblock {Pulsed contractions of an actin-myosin network drive apical
  constriction}.
\newblock {\em Nature}, 457(7228):495--499, 2009.

\bibitem{Tordesillas-Tobin-Cil-Alshibli-Behringer:2015}
A.~Tordesillas, S.T. Tobin, M.~Cil, K.~Alshibli, and R.P. Behringer.
\newblock {Network flow model of force transmission in unbonded and bonded
  granular media}.
\newblock {\em Phys. Rev. E: Stat. Nonlinear Soft Matter Phys.}, 91(6):062204,
  2015.

\bibitem{Kondic-Goullet-OHern-Kramar-Mischaikow-Behringer:2012}
L.~Kondic, A.~Goullet, C.S. O'Hern, M.~Kramar, K.~Mischaikow, and R.P.
  Behringer.
\newblock {Topology of force networks in compressed granular media}.
\newblock {\em EPL}, 97(5):54001, 2012.

\bibitem{Tewari-Schiemann-Durian-Knobler-Langer-Liu:1999}
S.~Tewari, D.~Schiemann, D.J. Durian, C.M. Knobler, S.A. Langer, and A.J. Liu.
\newblock Statistics of shear-induced rearrangements in a two-dimensional model
  foam.
\newblock {\em Phys. Rev. E}, 60(4):4385--4396, 1999.

\bibitem{Durian:1995}
D.J. Durian.
\newblock {Foam mechanics at the bubble scale}.
\newblock {\em Phys. Rev. Lett.}, 75(26):4780--4783, 1995.

\bibitem{Durian:1997}
D.J. Durian.
\newblock {Bubble-scale model of foam mechanics: Melting, nonlinear behavior,
  and avalanches}.
\newblock {\em Phys. Rev. E: Stat. Nonlinear Soft Matter Phys.},
  55(2):1739--1751, 1997.

\bibitem{Langer-Liu:1997}
S.A. Langer and A.J. Liu.
\newblock {Effect of random packing on stress relaxation in foam}.
\newblock {\em J. Phys. Chem. B}, 101(43):8667--8671, 1997.

\bibitem{Polyakov-He-Swan-Shaevitz-Kaschube-Wieschaus:2014}
O.~Polyakov, B.~He, M.~Swan, J.W. Shaevitz, M.~Kaschube, and E.~Wieschaus.
\newblock Passive mechanical forces control cell-shape change during drosophila
  ventral furrow formation.
\newblock {\em Biophys. J.}, 107:998--1010, 2014.

\bibitem{Hove-Koster-Forouhar-AcevedoBolton-Fraser-Gharib:2003}
J.R. Hove, R.W. K{\"o}ster, A.S. Forouhar, G.~Acevedo-Bolton, S.E. Fraser, and
  M.~Gharib.
\newblock Intracardiac fluid forces are an essential epigenetic factor for
  embryonic cardiogenesis.
\newblock {\em Nature}, 421:172--177, 2003.

\bibitem{Vermot-Forouhar-Liebling-Wu-Plummer-Gharib-Fraser:2009}
J.~Vermot, A.S. Forouhar, M.~Liebling, D.~Wu, D.~Plummer, M.~Gharib, and
  E.~Fraser, S.
\newblock {Reversing Blood Flows Act through klf2a to Ensure Normal
  Valvulogenesis in the Developing Heart}.
\newblock {\em PLoS Biol.}, 7(11):e1000246, 2009.

\bibitem{Bryngelson-Wolynes:1987}
J.D. Bryngelson and P.G. Wolynes.
\newblock {Spin-glasses and the statistical-mechanics of protein folding}.
\newblock {\em Proc. Natl. Acad. Sci. U.S.A.}, 84(21):7524--7528, 1987.

\bibitem{Li-Scheraga:1987}
Z.Q. Li and H.A. Scheraga.
\newblock {Monte-Carlo-minimization approach to the multiple-minima problem in
  protein folding}.
\newblock {\em Proc. Natl. Acad. Sci. U.S.A.}, 84(19):6611--6615, 1987.

\bibitem{Lau-Dill:1989}
K.F. Lau and K.A. Dill.
\newblock {A Lattice Statistical-Mechanics Model of the Conformational and
  Sequence-Spaces of Proteins}.
\newblock {\em Macromolecules}, 22(10):3986--3997, 1989.

\bibitem{Foffi-McCullagh-Lawlor-Zaccarelli-Dawson-Sciortino-Tartaglia-Pini-Stell:2002}
G.~Foffi, G.D. McCullagh, A.~Lawlor, E.~Zaccarelli, K.A. Dawson, F.~Sciortino,
  P.~Tartaglia, D.~Pini, and G.~Stell.
\newblock {Phase equilibria and glass transition in colloidal systems with
  short-ranged attractive interactions: Application to protein
  crystallization}.
\newblock {\em Phys. Rev. E: Stat. Nonlinear Soft Matter Phys.}, 65(3,
  1):031407, 2002.

\bibitem{Lee-Stell-Wang:2003}
C.L. Lee, G.~Stell, and J.~Wang.
\newblock {First-passage time distribution and non-Markovian diffusion dynamics
  of protein folding}.
\newblock {\em J. Chem. Phys.}, 118(2):959--968, 2003.

\bibitem{Leite-Onuchic-Stell-Wang:2004}
V.B.P. Leite, J.N. Onuchic, G.~Stell, and J.~Wang.
\newblock {Probing the kinetics of single molecule protein folding}.
\newblock {\em Biophys. J.}, 87(6):3633--3641, 2004.

\bibitem{Davis-Chin:2012}
L.~Davis and J.W. Chin.
\newblock {Designer proteins: applications of genetic code expansion in cell
  biology}.
\newblock {\em Nat. Rev. Mol. Cell Biol.}, 13(3):168--182, 2012.

\bibitem{Mandell-Lajoie-Mee-Takeuchi-Kuznetsov-Norville-Gregg-Stoddard-Church:2015}
D.J. Mandell, M.J. Lajoie, M.T. Mee, R.~Takeuchi, G.~Kuznetsov, J.E. Norville,
  C.J. Gregg, B.L. Stoddard, and G.M. Church.
\newblock {Biocontainment of genetically modified organisms by synthetic
  protein design}.
\newblock {\em Nature}, 518(7537):55+, 2015.

\bibitem{Procko-Hedman-Hamilton-Seetharaman-Fleishman-Su-Aramini-Kornhaber-Hunt-Tong-Montelione-Baker:2013}
E.~Procko, R.~Hedman, K.~Hamilton, J.~Seetharaman, S.J. Fleishman, M.~Su,
  J.~Aramini, G.~Kornhaber, J.F. Hunt, L.~Tong, G.T. Montelione, and D.~Baker.
\newblock {Computational Design of a Protein-Based Enzyme Inhibitor}.
\newblock {\em J. Mol. Biol.}, 425(18):3563--3575, 2013.

\bibitem{Jain-Inamdar-Padinhateeri:2015}
I.~Jain, M.M. Inamdar, and R.~Padinhateeri.
\newblock {Statistical Mechanics Provides Novel Insights into Microtubule
  Stability and Mechanism of Shrinkage}.
\newblock {\em PLoS Comput. Biol.}, 11(2):UNSP e1004099, 2015.

\bibitem{Heussinger-Bathe-Frey:2007}
C.~Heussinger, M.~Bathe, and E.~Frey.
\newblock {Statistical mechanics of semiflexible bundles of wormlike polymer
  chains}.
\newblock {\em Phys. Rev. Lett.}, 99(4):048101, 2007.

\bibitem{Mizuno-Tardin-Schmidt-MacKintosh:2007}
D.~Mizuno, C.~Tardin, C.F. Schmidt, and F.C. MacKintosh.
\newblock {Nonequilibrium mechanics of active cytoskeletal networks}.
\newblock {\em Science}, 315(5810):370--373, 2007.

\bibitem{Baumgart-Hammond-Sengupta-Hess-Holowka-Baird-Webb:2007}
T.~Baumgart, A.T. Hammond, P.~Sengupta, S.T. Hess, D.A. Holowka, B.A. Baird,
  and W.W. Webb.
\newblock {Large-scale fluid/fluid phase separation of proteins and lipids in
  giant plasma membrane vesicles}.
\newblock {\em Proc. Natl. Acad. Sci. U.S.A.}, 104(9):3165--3170, 2007.

\bibitem{Baumgart-Hess-Webb:2003}
T.~Baumgart, S.T. Hess, and W.W. Webb.
\newblock {Imaging coexisting fluid domains in biomembrane models coupling
  curvature and line tension}.
\newblock {\em Nature}, 425(6960):821--824, 2003.

\bibitem{Lingwood-Simons:2010}
D.~Lingwood and K.~Simons.
\newblock {Lipid Rafts As a Membrane-Organizing Principle}.
\newblock {\em Science}, 327(5961):46--50, 2010.

\bibitem{Morin-Daneman-Zavortink-Chia:2001}
X.~Morin, R.~Daneman, M.~Zavortink, and W.~Chia.
\newblock {A protein trap strategy to detect GFP-tagged proteins expressed from
  their endogenous loci in Drosophila}.
\newblock {\em Proc. Natl. Acad. Sci. U.S.A.}, 98(26):15050--15055, 2001.

\bibitem{Cavey-Lecuit:2008}
M.~Cavey and T.~Lecuit.
\newblock {\em Imaging cellular and molecular dynamics in live embryos using
  fluorescent proteins}, volume 420 of {\em Methods in Molecular Biology}.
\newblock Humana Press Inc., Totowa, NJ, 2008.

\bibitem{Spencer-Siddiqui-Thomas:2015}
A.K. Spencer, B.A. Siddiqui, and J.H. Thomas.
\newblock Cell shape change and invagination of the cephalic furrow involves
  reorganization of f-actin.
\newblock {\em Dev. Biol.}, 402:192--207, 2015.

\bibitem{Figard-Sokac:2011}
L.~Figard and A.M. Sokac.
\newblock {Imaging Cell Shape Change in Living Drosophila Embryos}.
\newblock {\em J. Vis. Exp.}, 49, 2011.

\bibitem{Schneider-Rasband-Eliceiri:2012}
C.A. Schneider, W.S. Rasband, and K.W. Eliceiri.
\newblock {NIH Image to ImageJ: 25 years of image analysis}.
\newblock {\em Nat. Methods.}, 9(7):671--675, 2012.

\bibitem{gao06}
G.-J. Gao, J.~Blawzdziewicz, and C.S. O'Hern.
\newblock {Frequency distribution of mechanically stable disk packings}.
\newblock {\em Phys. Rev. E:}, 74:061304, 2006.

\end{thebibliography}

\end{document}